\documentclass[preprint,12pt]{elsarticle}

\usepackage{amssymb}
\usepackage{amsmath}
\usepackage{booktabs}
\usepackage{multirow}
\usepackage{longtable}
\usepackage{natbib}
\usepackage{caption}
\journal{Design and Artificial Intelligence}

\begin{document}

\begin{frontmatter}



\title{Exploring Micro Accidents and Driver Responses in Automated Driving: Insights from Real-world Videos} 


\author{Wei Xiang$^a$, Chuyue Zhang$^a$, Jie Yan$^{a*}$} 

\affiliation{organization={International Design Institute, Zhejiang University},
            city={Hangzhou},
            postcode={310058}, 
            country={China}}

\begin{abstract}
Automated driving in level 3 autonomy has been adopted by multiple companies such as Tesla and BMW, alleviating the burden on drivers while unveiling new complexities. This article focused on the under-explored territory of micro accidents during automated driving, characterized as not fatal but abnormal aberrations such as abrupt deceleration and snake driving. These micro accidents are basic yet pervasive events that might results in more severe accidents. Through collecting a comprehensive dataset of user generated video recording such micro accidents in natural driving scenarios, this article locates key variables pertaining to environments and autonomous agents using machine learning methods. Subsequently, crowdsourcing method provides insights into human risk perceptions and reactions to these micro accidents. This article thus describes features of safety critical scenarios other than crashes and fatal accidents, informing and potentially advancing the design of automated driving systems.
\end{abstract}

\begin{graphicalabstract}
\end{graphicalabstract}

\begin{highlights}
\item Two-thirds of drivers failed to recognize risky situations before micro accidents occurred
\item Natural driving videos revealed critical gaps in human-machine collaboration patterns
\item Crowdsourcing revealed drivers misjudge risk by focusing on severity over probability
\item Rural roads and complex intersections pose greatest challenges for autonomous systems
\end{highlights}


\end{frontmatter}



\section{Introduction}
The introduction of level 3 driving autonomy, as observed in companies like Tesla, Xiaopeng, Uber, etc., has relieved drivers from repetitive and tedious tasks \citep{SAE2021Taxonomy}, while also brings new types of challenges and accidents. Automated agents might not perform well in specific driving contexts; drivers might fail to monitor the automated driving agents and cope with sudden risky events. Facing these new features, the causes of accidents attracted researchers’ interest over an extended period \citep{Chu2023Work}.

Existing studies have employed settings such as driving simulators, Wizard-of-Oz experiments, and case analysis to locate key variables affecting safe driving. They have validated the effects of multiple variables involving velocity, alert modalities, and drivers’ risk perceptions \citep{Kunze2018Augmented, Techer2019Anger}. These findings informed the design of warning visualization and proactive service during driving \citep{Colley2021How, Kim2020Interruptibility}. However, the studies primarily focused on fatal accidents such as crashes \citep{Banks2018Driver}, often overlooking lower risk, yet still abnormal events. We called these safety critical events with low risk as micro accidents, such as sharp braking, route deviation, and snakelike motion. Though not immediately hazardous, these micro accidents occurred frequently and served as the precursor of more severe accidents, making them a significant category of abnormal events that required researchers’ attention and devotion. Remarkably, scant studies have addressed these micro accidents in natural driving contexts, leaving a knowledge gap regarding their relevant variables and drivers’ responses.

This article noticed this research gap and focused on real-world micro accidents in Level 3 automated driving. Natural driving data have irreplaceable value and have been addressed by multiple researchers. With the wide application of automated driving in recent years, a vast number of automated driving videos have been generated and uploaded by drivers, providing rich data of micro accidents in natural scenarios. This article conducted a three-step analysis on micro accidents in natural driving scenarios. First, we collected over 400 videos online and annotated the features of micro accidents. Second, we employed machine learning methods in conjunction with SHapley Additive exPlanations (SHAP) to identify variables correlated with diverse micro accidents. Third, to compensate the absence of direct observational data of drivers’ faces and responses, we utilized crowdsourcing methods involving over 400 participants, to analyze participants’ risk perception and reaction to these micro accidents. This integrative approach enriched our comprehension on the variables and human responses associated with micro accidents.

This article has the following contributions:

1.	This article revealed key variables correlated with micro accidents, involving both environmental variables and variables of autonomous agents. These variables extended our understanding of safety critical but not fatal incidents beyond crashes and catastrophic accidents, thereby facilitating micro accident identification and prediction.

2.	This article explored drivers’ perception around micro accidents, revealing a potential difference between perceived and actual risk. This discrepancy might explain drivers’ inappropriate behavior in such contexts and inspire further design of automated driving systems.

3.	This article collected a vast number of user-generated videos recording real driving scenarios. It employed machine learning and crowdsourcing to analyze the perceptual and reactive features of drivers and autonomous agents. The data and method enriched the empirical base for future research.
\section{Related Work}
\subsection{Variables Affecting Driving Safety}
Automated driving at level 3 constructs a collaborative relationship between drivers and autonomous agents. Autonomous agents drive; drivers adopt a supervisory role and intervene whenever needed \citep{Cabrall2019How, Huysduynen2018Why}. This new role challenges drivers’ perceptions and behavior patterns \citep{Banks2018Is, Liu2021Blame}. For instance, researchers have reported that drivers had a poor mental model of automated driving, were reluctant in hazard responses and take over requests \citep{Janssen2019Interrupted}, and struggled to calibrate optimal cognitive workload (either too relaxed or too intense) \citep{Merriman2021Challenges}. These challenges affected the efficiency and safety of automated driving. 

The reasons why drivers and cars fall into risky scenarios is multifaceted, incorporating variables from road conditions to system variables. For example, automated vehicles performed well on the highway while still struggling in urban areas. Interfaces and interactions, which incorporated rich functions, alert events while also distracting drivers \citep{Grahn2020Impacts, Hafizi2023In}. The warning improves perceived safety irrespective of the visualization concepts that automated driving system employ \citep{Colley2021How}. Besides, incorporating warnings in continuous music through matching warning levels with music pitches also worked well \citep{Chen2021Manipulating}. However, the interfaces also supported in-vehicle multitasking and brought varied levels of driver interruptibility \citep{Kim2020Interruptibility}. 

Human factors, such as risk perception and expectations also affected driving safety. Drivers adjusted their driving behavior according to their anticipated event probabilities and potential consequences \citep{Kolekar2020Human}. As evidenced in both simulation-based studies and real world collision analyses, when drivers did not expect an accident, they failed to avoid crashes \citep{Ba2016How}, and made late interventions \citep{Stanton2019Models}. On the contrary, drivers knowing a high possibility of hazard events paid less attention to secondary tasks \citep{Lin2019Exploring}. Distinct variables, such as monetary incentives, the number of vehicles, distance to action, speed, weather, position, and driving route, further complicate driving choices \citep{Clark2019Conditionally, Cohen2018Money}. There is a divergence between drivers’ behavior preference and their self-perceived optimal behavior. Drivers chose not to be intervened in most scenarios, while they thought others should be intervened in the same scenario \citep{Maurer2019Playing}. 

Given these intricate factors affecting automated driving, key variables associated with micro-accidents could not be directly summarized from studies on normal driving and accident scenarios. When, where, and how micro accidents happen in automated driving remain largely unexplored.

\subsection{Data on Automated Driving}
Although driving simulators supported a large proportion of studies because of their flexibility in scenario setting and safety, they cannot fully replicate the nuances of real-world driving experiences. Winter et al. suggested a shift to realistic scenarios while reviewing the takeover studies \citep{Winter2021Is}. Interviews of drivers across multiple domains involving maritime, rail, and aviation have provided effective guidelines \citep{Trosterer2017What}. Guided by these advocates, Reimer and Fridman launched MIT Advanced Vehicle Technology (MIT-AVT) dataset collection and recorded large-scale real-world driving data \citep{Fridman2019MIT}. These data are mostly normative driving scenarios, supporting training of autonomous algorithms and analyses of drivers’ behavior \citep{Al2023Keep, Ding2020MIT, Morando2020Driver}. Similarly, Koch et al. examined drivers’ reaction to interventions during driving on public roads \citep{Koch2021When}. 

Nevertheless, most studies analyzed limited reports of crashes \citep{Merriman2021What}, and employed qualitative methods as an initial exploration of automated driving problems \citep{Brown2023Halting}. The data and quantitative studies are scarce for abnormal driving scenarios, with a notable absence of first-person observational data that display conditions of micro accidents and driver’s responses. This impedes the comprehensive analysis of such critical yet underexplored events.

\subsection{Analysis Methods of Automated Driving}
Several analytical methods, such as statistical models and machine learning \citep{Lord2010statistical}, have supported the analyses of driving data. Support vector machine (SVM) has been particularly efficacious to address complex, large, and multidimensional spatial data in crash prediction \citep{Dong2015Support}, and achieved high accuracy \citep{Parsa2019Real}. Other algorithms like k-nearest neighbors, regression trees, feedforward neural networks, and deep learning models have also been applied to detect accidents \citep{Ozbayoglu2016real, Parsa2019Applying}. These models achieved a high prediction accuracy yet hard to locate the influential variables. 

Shapley Additive exPlanations (SHAP), which interprets the model output based on cooperative game theory, provides a way to explain influential variables within machine learning model.\citet{Zhou2022Using} obtained the variable importance ranking using lightGBM and SHAP to select influential variables, and achieved enhanced performance in the prediction of situational awareness. \citet{Ayoub2021Modeling} also employed SHAP and XGBoost model to refine the prediction performance of trust. SHAP explained the results of convolutional neural network model and analyzed the factors that increase the collision risk of intersection \citep{Hu2020Efficient}. \citet{Wen2021Quantifying} explained the factors affecting collision types with lightGBM and SHAP. \citet{Parsa2020Toward} also used XGBoost model to detect accidents, and used SHAP to explain and analyze the importance of variables. These precedent studies affirmed the utility of employing machine learning models and SHAP for video data analysis in our context.

Generally, our study diverges from existing studies by specifically focusing on the underexamined micro accidents in level 3 automated driving. These micro accidents are significant as they are the basic while ubiquitous abnormal events that requires drivers’ attention and response. To analyze the scarce data of micro accident in natural scenes of automated driving, we collected user updated video on multiple online platforms. The key variables affecting micro accidents were located using SHAP methods. This aligns with qualitative research of \citet{Brown2017Trouble}, who collected online videos to locate the challenges associated with Autopilots.

\section{Methodology}
\subsection{Overall Structure}
This article conducted real scenario video analyses and crowdsourcing survey to identify the features of micro accidents in level 3 automated driving. We collected automated driving videos and annotated the driving context, the behavior of autonomous agents, and the type of micro accidents. The variables correlated with micro accidents were explored employing machine learning classification models and SHAP. 

Given that these videos rarely capture drivers’ perception of riskiness, a subsequent crowdsourcing study was conducted to explore participants’ perceptions and responses to micro accidents. This addressed two questions: 1) When participants watched pre-accident driving scenarios, how did they feel and evaluate the riskiness? 2) When participants watched the micro accident, how did they response? Researchers have examined the validity of crowdsourcing participants on experimental study \citep{Thomas2017Validity}. Also, the video, as explained by Transportation Theory, provided a media to make participants cognitively and emotionally involved in the driving scenarios as long as participants have prior knowledge \citep{Green2004Transportation}. This allows for participant feedback that closely approximates the reactions of the actual drivers in the videos. 

Combining video analyses and crowdsourcing, this study hopefully revealed the variables invoking micro accidents and illuminate drivers’ perceptions of these accidents. The following subsections described procedures of video collection, video analysis, and crowdsourcing approach.

\subsection{Video Collection and Annotation}
The video collection aims to capture a wide spectrum of real scenarios specifically at SAE level 3. Videos were collected from multiple platforms, including Youtube, Twitter, Douyin, Bilibili, and Youku, ensuring geographical and cultural diversity in the data. A set of searching keywords related to micro accidents are listed in Table \ref{tab:keywords}. These keywords involved synonyms and related terms to accidents, challenging conditions like traffic jam, and actions often associated with emergency situations such as takeover.

Inclusion criteria specified that videos shall be shot from a driver’s first-person perspective, capturing the steering wheel and central control screen, and the road conditions through windshield. This ensured that video data provide information on the driving environment, the interfaces of the autonomous systems, and any drivers’ activities. The video also had to confirm the SAE level 3 of automated driving, during which the automated system drives and the driver only monitors, and also verified through checking the brand of vehicles. It is worth noting that we did not collect non-user generated videos, involving advertisement, and driving system exhibition videos made by companies. This criterion ensured that the dataset comprised user experience during real world automated driving.

The searching process followed a structured approach: 

1.	On each platform, searched a keyword, looked up the first ten pages and collected videos that met the inclusion criteria. Keywords were adapted and combined to maximize the range of scenarios captured (for example, automated driving, automated driving plus avoid, and Tesla plus slow).

2.	If there were not any related videos on the first ten pages, the search for that keyword was considered complete on that platform. If related videos were found on a page, the search continues for another five pages. This manner continued until no further related videos were found.

The threshold of ten and five pages was empirically determined. Repeated searches across multiple keywords and platforms ensured that a sufficiently large and representative sample of videos was collected. Still, a larger sample could offer even more comprehensive insights. 

\begin{table}[htbp]
\centering
\caption{Keywords Searched in Video Platforms}
\label{tab:keywords}
\begin{tabular}{lll}
\toprule
Website & Keywords &   \\
\midrule
Twitter, Youtube & Automated driving & Change lane \\
Douyin, Bilibili, Youku* & Autonomous driving & Lane change \\
                 & Auto pilot       & Traffic jam \\
                 & Tesla            & Slow \\
                 & Takeover         & Avoid \\
                 & Handover         & Fail \\
                 & Emergency        & Wrong \\
                 & Accident         & Evasive \\
                 & Crash            & Obstacle \\
                 & Collision        & Other drivers \\
                 & Pedestrian       & Dodge \\
\bottomrule
\end{tabular}

\smallskip
\footnotesize
* Douyin, Bilibili, and Youku are Chinese websites; we translated the keywords into Chinese while searching in these three websites.
\end{table}

Through a manual screening process, an initial pool of 490 videos were collected. Then, we manually reviewed the videos, excluded repetitive videos and videos that did not contain clear road conditions, finally built a dataset of 277 unique clips, each focused on a singular micro accident. These video clips have a resolution of 1280 pixels * 720 pixels, displayed a temporal window extending from approximately 20s before the micro accident to 10 seconds after the micro accident, lasting a total duration of 30 seconds. It is worth noting that we excluded a large number of normal driving videos (for example, a well-functioned Autopilot on a highway). Then, these video clips were categorized into four types of micro accidents referring to the incident type in prior natural driving studies.

\begin{longtable}{p{2.5cm}p{4.5cm}p{7cm}}
\caption{Variables of Environments and Autonomous Agents}\label{tab:table2} \\
\hline
Category & Variable value & Description \\
\hline
\endfirsthead

\multicolumn{3}{c}%
{{\tablename\ \thetable{} (continued from previous page)}} \\
\hline
Category & Variable value & Description \\
\hline
\endhead

\hline \multicolumn{3}{|r|}{{Continued on next page}} \\
\endfoot

\hline
\endlastfoot

Dependent Variable & Micro\_Accident (0,1,2,3) & LaneA = 0; \\
& & ChangeB = 1; \\
& & ObstacleC = 2; \\
& & ViolationD = 3; \\
\hline

Environment & Weather (0,1,2) & Sunny day = 0; \\
& & Small rain, fog, snow, night with light = 1; \\
& & Heavy rain, fog, snow, dark night, vision impaired = 2; \\
\cline{2-3}
& Road\_Type (0,1,2,3,4,5) & Garages and parking lots = 0; Rural roads = 1; Urban roads = 2; \\
& & High speed = 3; Elevated = 4; Mountain roads = 5; \\
\cline{2-3}
& Number\_of\_Cars (0,1,2,3) & No other vehicles = 0; \\
& & Small (1-3) = 1; \\
& & Medium (>3) = 2; \\
& & Congestion = 3; \\
\hline

Errors & Computer\_Vision (0,1,2,3) & No abnormality = 0; \\
& & Abnormal recognition of objects = 1; \\
& & Abnormal lane line recognition = 2; \\
& & Camera obscured = 3; \\
\cline{2-3}
& Congestion (0,1) & None = 0; \\
& & Blocked by other vehicles = 1; \\
\cline{2-3}
& Lane\_Change (0,1) & None = 0; \\
& & Failure to change lane or turn when it could = 1; \\
\cline{2-3}
& Off\_Center (0,1) & None = 0; \\
& & Off-center lane = 1; \\
\cline{2-3}
& Route\_Planning (0,1) & None = 0; \\
& & Plan a wrong route (ignore ring road, auxiliary road, etc.) = 1; \\
\cline{2-3}
& Acceleration (0,1) & None = 0; \\
& & Sharp acceleration and deceleration in normal driving = 1; \\
\cline{2-3}
& Sudden\_Obstacles (0,1) & None = 0; \\
& & Sudden appearance of obstacle = 1; \\
\hline

Vehicle action & Vehicle\_action (0,1,2,3,4,5) & Will meet an intersection = 0; \\
& & Drive in the current lane and remain straight = 1; \\
& & Drive in the current lane and will enter a curve = 2; \\
& & Encounter complex intersections such as traffic circles, off-ramps, ramps, etc. = 3; \\
& & Change lanes = 4; \\
& & Reverse = 5; \\
\hline

Intervention Action & Intervention\_Rationality (0,1,2) & No intervention = 0; \\
& & Unreasonable or unnecessary interventions = 1; \\
& & Reasonable interventions = 2; \\
\cline{2-3}
& Send\_Reminder (0,1) & None = 0; \\
& & Warning sound = 1; \\
\cline{2-3}
& Deceleration\_or\_ \newline Emergency\_Braking (0,1) & None = 0; \\
& & Slow down or brake sharply = 1; \\
\cline{2-3}
& Lane\_Changing\_or\_ \newline Avoidance (0,1) & None = 0; \\
& & Lane change or avoidance = 1; \\

\end{longtable}

LaneA – unstable lane keeping. For example, erratic lateral movements in the lane or drive across the line.

ChangeB - risky multi-lane action. For example, wrong lane changes or incorrect route selection at forks.

ObstacleC - emergent braking. For example, the system failed to recognize an obstacle, leading to a near collision sharp braking.

ViolationD – traffic rule violation. For example, red light running and contraflow driving.

The 277 video clips include 85 clips of LaneA (30.69\%), 118 (42.60\%) clips of ChangeB, 55 (19.86\%) clips of ObstacleC, and 19 (6.86\%) clips of ViolationD.

Subsequent to the categorization, these video clips were then annotated according to the taxonomy displayed in Table \ref{tab:table2}. Given that each video clip contained a singular micro accident, annotations were focused on the features of the driving scenarios within which micro accident happened. The variables cover a broad range of factors during driving, including environmental conditions, autonomous agent errors, vehicle actions, and intervention action executed by the autonomous agents. Compared with variables annotated in multiple prior natural driving dataset \citep{Fridman2019MIT, Van2011Towards}, variables related to driver features was not included due to the limitation of the video dataset. Besides, the annotation scheme in table 2 added errors and intervention action to emphasize the micro accidents. Variable identification mainly relied on recognizing the road conditions, while attributes like Computer\_Vision, Route\_Planning, and Send\_reminder were recognized from the information on central control screen. Only intervention signals shown on the interface would we annotate the variable in the intervention action.

Two researchers first marked 50 videos, discussed the differences, and unified the labeling standards (average kappa=0.95 for the 15 variables). Subsequently, we recruited four participants to annotate the remaining videos, which are college students with driver license. Each tagger first joined in a training process and then annotated videos. Then, the two researchers reviewed all the annotation to ensure the accuracy and reliability of the dataset.

\subsection{Video Analysis}

\subsubsection{Predictive Classification of Micro Accidents}
This study chose XGBoost classifier to predict the micro accidents \citep{Chen2016XGBoost}. XGBoost algorithm is an integrated learning method with classification and regression tree (CART) as the base classifier \citep{Klemelii2000Classification}. XGBoost enhances the robustness of the model by introducing regular terms and column sampling, and also expedites the speed of the model through parallelized tree splitting. In addition, XGBoost is inherently free from multicollinearity. Even if there are positively or negatively correlated variables in the dataset, XGBoost seldom produce distorted or misleading results. This feature is beneficial for our subsequent analysis of the influence of variables.

Hyperparameter adjustment is one of the key steps in the process of model training and model fitting. Employing a grid search method, we instantiated a given model for each combination in the hyperparameter combination list, performed ten rounds of cross-validation, and the hyperparameter combination with the highest average score was taken as the best choice and returned to the model object. The best hyperparameter combinations for the final model were: learning\_rate=0.07, n\_estimators=30, max\_depth=4, min\_child\_weight = 4, gamma=0.2, subsample=0.7, colsample\_btree=0.8, reg\_alpha=2, and reg\_lambda=1. 

To ensure an accurate estimation of Micro-Accidents, we used a 10-fold cross-validation method to calculate Accuracy, Precision, Recall, and Macro-F1. We also compared the performance of XGBoost with four other popular machine learning models (see Table \ref{tab:table3}). The results showed the appropriateness of using XGBoost model in this study.

\begin{table}[htbp]
\centering
\caption{Performance Comparison Among Multiple Models}
\label{tab:table3}
\begin{tabular}{ccccc}
\toprule
Model & Accuracy & Precision & Recall & Macro-F1\\
\midrule
Decision Tree & 70.00\% & 54.94\% & 53.65\% & 51.59\& \\
Random Forest &	71.52\% & 54.99\% &	56.56\%	& 54.96\% \\
LightGBM & 73.60\% & 55.75\% & \textbf{57.94\%} & 56.05\% \\
KNN	& 65.67\% &	49.28\% & 53.21\% & 50.49\% \\
XGBoost & \textbf{73.68\%} & \textbf{56.22\%} & 57.80\% & \textbf{56.10\%} \\
\bottomrule
\end{tabular}
\end{table}

\subsubsection{Explain XGBoost Model Using SHAP}
This study used SHAP to interpret the output of XGBoost models \citep{Lundberg2017unified}. SHAP constructed an additive explanatory model where all features (i.e., predictor variable in XGBoost) are considered contributors. For each prediction sample, the model generated a prediction value, and the SHAP value was the value assigned to each variable in that sample. We used SHAP to analyze the effect of variables on the four categories of micro accidents. The SHAP summary plot showed the impact of each predictor variable on micro accidents. Then, the SHAP main effect plot displayed the relationship between the values of variables and the SHAP value in detail.

\subsection{Crowdsourcing Study}
The crowdsourcing experiment explored drivers’ perceptions of micro accidents. We randomly chose 40 videos from the video clips and made two versions for each video: the video without micro accidents and the original video with micro accidents, producing a total of 80 videos. Video without micro accidents was made by deleting the micro accident from 0.4 seconds just before it happened and did not show the following emergent events. The 0.4 seconds were decided referring to people’s perception time (around 0.2 seconds) plus judgment time (around 0.2 seconds). The UN regulation on Automated Lane Keeping systems (ALKS) also proposed 0.4 seconds as the time a driver needed to perceive risk. Therefore, it is reasonable to assume that drivers should recognize possible micro accidents 0.4 seconds before it happened. 

The sample size of 40 was chosen considering the variety of videos and the scale of crowdsourcing study. Participants were recruited from Amazon Mechanical Turk and remunerated 0.5 dollars for their participation. All participants were Amazon Turk master workers, had an approval rate greater than 95, and had a driver license and familiarity with automated driving. 

During the crowdsourcing experiment, participants first answered their experience of driving and familiarity with automated driving. Then, they read a paragraph introducing the survey, watched a video and answered the following questions. The paragraph described the video content: “The video shows a scenario of automated driving, where the automated system drives and the driver monitors at the level of SAE level 3. You will act as the driver to monitor the automated system and judge the riskiness of the video.” Then, it asked participants to watch the first perspective video and answer questions. All statements were measured using a 7-point Likert scale. A typical video last around 30 seconds, therefore the whole process lasted around 3-5 minutes. The time was set for two reasons. First, crowdsourcing studies work well for short time surveys, so each survey only involved one video and participants would not be interrupted by the content in other videos. Second, this elicited a first impression reaction similar to the driver in the video rather than deliberative assessments.

\begin{enumerate}
    \item For participants watching videos without micro accidents, they needed to decide if there would be an accident using a 7-point Likert scale, and if so, what was the micro accident. 7 indicated high possibility and high riskiness, and 1 indicated low possibility and low riskiness. The queries were as follows:
    \begin{enumerate}
        \item Please rate the probability of any emergent events in this situation. (7-point Likert scale)
        \item Please rate the dangerousness of the emergent events in this situation if some emergency would happen. (7-point Likert scale)
        \item Please rate the riskiness of this driving situation. (7-point Likert scale)
        \item If you think any possible emergent event might happen in the video, please briefly describe it. If not, please describe the weather and the number of roads in the video. (open ended text)
    \end{enumerate}
    \item For participants watching original videos, participants answered:
    \begin{enumerate}
        \item Please rate the riskiness of this driving situation. (7-point Likert scale)
        \item Who do you owe the emergent events to? Please rate drivers’ responsibility for this emergency. (7-point Likert scale)
        \item Who do you owe the emergent events to? Please rate the responsibility of autonomous agents for this emergency. (7-point Likert scale)
        \item How this driving situation can be improved if you are the driver in the video? (e.g. autonomous agents, drivers, etc.) (open ended text)
    \end{enumerate}
\end{enumerate}

This experiment collected a total of 480 surveys, indicating 6 ratings for each video. The ones that did not complete the survey, and who answered wrong weather and number of roads in the video were rejected, resulting in a total of 421 valid surveys, including 189 surveys for video without micro accidents and 232 surveys for original videos. All participants except 11 have more than three years of driving experience, among which 215 have more than 10 years of driving experience. Then, the ratings of the same video were averaged for further analysis. 

\section{Results}
The results involve two parts. The first part reports results from SHAP analysis and the second part reports participants’ perception and response to the micro accidents.
\subsection{Variables that Invoke Micro Accidents}
SHAP summary plots depict a comprehensive relationship between 15 SHAP variables and four micro accidents (Figure \ref{fig:figure1}).  The variables are prioritized based on their average absolute shape-value, serving as an index for their respective importance. Vehicle\_action emerges as an important variable in predicting micro accidents, namely the driving contexts (current lane or intersections) affected the frequency of micro accidents. The SHAP summary plots displayed the varied effect of variables on the four micro accidents. Following sections used the SHAP main effect plot to explain the effect of each variable in detail.
\subsubsection{Environmental Variables}
Interestingly and unexpectedly, the weather has a significant negative drive for type ChangeB micro accidents, indicating that bad weather was not the reason for such events. Only in Figure 1a and Figure 1c can we see that weather has a slight positive effect (taking a non-zero value) on LaneA and ObstacleC. This implies that bad weather conditions might decrease lane change actions while slightly increase the probability of accidents, consistent with existing literature \citep{Parsa2020Toward}.

\textbf{Different road types had varied effects on the four types of micro accidents} (variable Road\_Type, Figure \ref{fig:figure2}). LaneA accidents were prone to the type of road (Figure \ref{fig:figure2}a). Rural roads have more ChangeB accidents, such as anomalies during lane changes and making turns (Figure \ref{fig:figure2}b). ViolationD accidents were more likely to occur on urban roads (Figure \ref{fig:figure2}c). Remarkably, the road types variable had not a significant effect on ObstacleC, indicating no direct relationship between the type of road and emergent braking or obstacle avoidance behaviors.

\begin{figure}
    \centering
    \includegraphics[width=\linewidth]{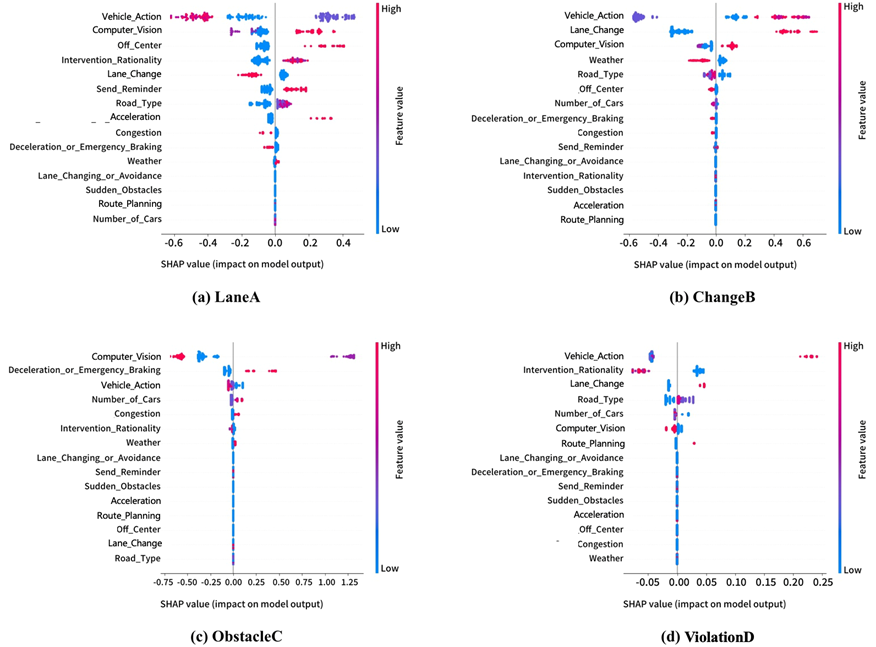}
    \caption{SHAP summary plot. Each point represents a sample, and the aggregation of points reflects the distribution of the samples. The horizontal axis represents the SHAP value relative to the baseline value. A positive SHAP value indicates that the variable has a positive driving force for the determination of micro accidents, i.e., increases the probability. A negative SHAP value indicates that the variable has a negative driving force for the determination of micro accidents, i.e., decreases the probability.}
    \label{fig:figure1}
\end{figure}

\begin{figure}
    \centering
    \includegraphics[width=\linewidth]{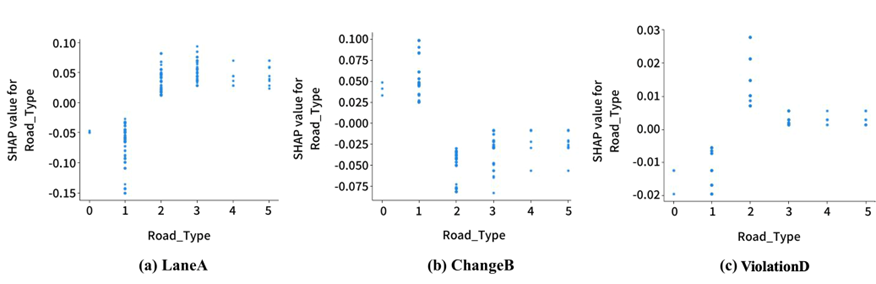}
    \caption{SHAP Main Effect Plot of Road Type}
    \label{fig:figure2}
\end{figure}

\begin{figure}
    \centering
    \includegraphics[width=0.5\linewidth]{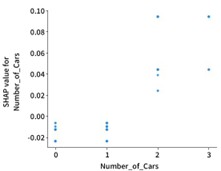}
    \caption{SHAP Main Effect Plot of Number\_of\_Cars (ObstacleC)}
    \label{fig:figure3}
\end{figure}

\textbf{Vehicluar density, represented by the variable “Number\_of\_Cars”, are not correlated with LaneA, ChangeB, and ViolationD} (Figure \ref{fig:figure1}a, Figure  \ref{fig:figure1}b, and Figure \ref{fig:figure1}d). The probability of ObstacleC increased when there are many vehicles or congestion on the road (Figure \ref{fig:figure1}c and Figure \ref{fig:figure3}). 

\subsubsection{Variables Correlated with Autonomous Agents}
\begin{figure}
    \centering
    \includegraphics[width=0.9\linewidth]{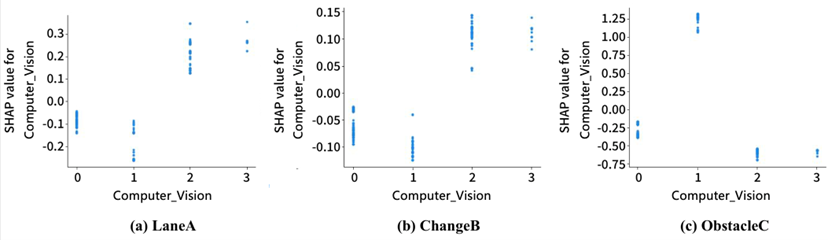}
    \caption{SHAP Main Effect Plot of Computer Vision}
    \label{fig:figure4}
\end{figure}

On account of autonomous agents, the importance of computer vision problems is in the top three for LaneA, ChangeB and ObstacleC, among which ObstacleC were the most affected (variable CV\_Error, Figure \ref{fig:figure4}). LaneA and ChangeB accidents were associated with unclear lane delineation and camera blocking (Figure \ref{fig:figure4}a, Figure \ref{fig:figure4}b). ObstacleC accidents were correlated with the incorrect recognition of people, vehicles, obstacles, and road signs on the road (Figure \ref{fig:figure4}c).

Other than computer vision, the variable of errors such as wrong route planning and inappropriate lane change contribute to the occurrence of LaneA, ChangeB, and ViolationD. Inappropriate driving behavior, for example, choosing a wrong route when another car is nearby, would increase the probability of micro accidents such as sharp breaking or violation.

\begin{figure}
    \centering
    \includegraphics[width=0.9\linewidth]{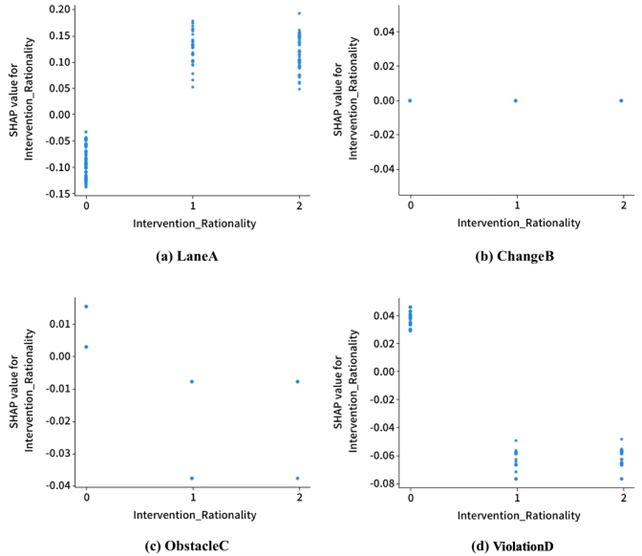}
    \caption{SHAP Main Effect Plot of Intervention\_Rationality}
    \label{fig:figure5}
\end{figure}

For LaneA micro accidents, autonomous agents performed well in alerting through audible cues (variable Send\_Reminder in Figure \ref{fig:figure1}a). Because the intervention was mainly warning sounds, human intervention was still needed to control the risk. In Obstacle C scenarios, the autonomous agents decelerated or braked (variable Deceleration\_or\_Emergency\_Braking in Figure \ref{fig:figure1}c), while their interventions are not consistently reasonable (Figure \ref{fig:figure5}c). Regarding ChangeB and Violation D type of micro accidents, autonomous agents currently lack a coherent intervention strategy and requires driver intervention (Figure \ref{fig:figure5}b and Figure \ref{fig:figure5}d). This raise questions about the overall utility and readiness of autonomous interventions in these scenarios.

\subsubsection{Variables Associated with Micro Accidents}
The results above proposed the salient variables that easily induce micro accidents (Table \ref{tab:table4}). Road related variables affect the performance of automated driving, particularly bring challenges in computer vision and road planning for rural areas and complex intersections. These scenarios often involved roads with construction fences, sharp turns, and highway fork. Besides, autonomous agents coped well with some of the static obstacles while could not handle suddenly appeared objects, which were also hard for human drivers.

\begin{table}[htbp]
\centering
\caption{Variables that Might Invoke Micro Accidents, the Intervention Behavior of Autonomous Agent and Driver}
\label{tab:table4}
\begin{tabular}{p{2cm}p{3.5cm}p{3.5cm}p{3.5cm}}
\toprule
Micro accidents & Variables of Environment and Autonomous agents & Interventions & Examples\\
\midrule
LaneA & Vehicle action, CV Error, Lane Change or Turning, Road type & Audible and visual reminder & Unclear lane delineation and camera blocking \\
ChangeB & Vehicle action, Weather, Road type, CV Error, Lane Change or Turning & Agent did not intervene, Driver intervened & Rural roads, Unclear lane delineation and camera blocking, Complex intersections, Highway fork \\
ObstacleC & Vehicle action, Acceleration, Number of cars, CV Error & Deceleration or braking & Many cars, Incorrect recognition of objects and road signs \\
ViolationD	& Vehicle action, Road type, Lane Change or Turning & None intervention & Urban roads \\
\bottomrule
\end{tabular}
\end{table}

\subsection{Perceptions and Responses to Micro Accidents in Crowdsourcing Study}
This sub section reports results from crowdsourcing study. Here the perception refers to perception of the driver of the ego vehicle, not drivers of other cars that observe automated cars in their surroundings.
\subsubsection{Perceived Riskiness of the Scenario in Video Clips}
The perceived riskiness of scenarios rated by participants increased after they saw the micro accidents (t=-2.908, p=0.006). Participants rated an average score of 4.01 for video ended 0.4 seconds before micro accidents and 4.61 for original videos. The relation between riskiness and the possibility and dangerousness of accidents was analyzed using regression analysis. The riskiness of the scenario mainly changed with participants’ expectations of the dangerousness of possible accidents (beta=0.770) rather than the possibility of possible accidents (beta=0.299). 

\subsubsection{Variables Affecting Participants’ Perception of Videos without Micro Accidents}
The relation between participants’ perceptions of the videos and the variable of video clips was analyzed using ANOVA. For videos ended before micro accidents, errors of computer vision affected the possibility of micro accidents (F=3.457, p=0.028), and dangerousness of micro accidents (F=3.245, p=0.033). The one that displays computer vision problems got higher possibility ratings and higher dangerousness ratings (Table \ref{tab:table5}). 

\begin{table}[htbp]
\centering
\caption{Ratings Affected by Errors of Computer Vision (CV\_Error)}
\label{tab:table5}
\begin{tabular}{lcc}
\toprule
    & Possibility & Dangerousness\\
\midrule
CV\_Error (0) No abnormality & 3.94 & 3.76 \\
CV\_Error (1) Abnormal recognition of objects & 4.03 & 3.80 \\
CV\_Error (2) Abnormal lane line recognition & 4.78 & 4.05 \\
\bottomrule
\end{tabular}
\end{table}

\subsubsection{Variables Affecting Participants’ Perception of Original Videos}
The original videos displayed risky driving scenarios during which some of the drivers intervened. Therefore, we also coded drivers’ intervention as Driver\_Intervention (0: did not intervene, 1: intervened) in original videos besides the variables in Table 2, and explored participants’ perception of riskiness and responsibility. For participants who watch the original videos, the riskiness of micro accidents is correlated with Intervention\_Rationality, Micro\_Accident, and Driver\_Intervention (Intervention\_Rationality: F=8.350, p=0.004; Micro\_Accident: F=4.530, p=0.020; Driver\_Intervention: F=8.801, p=0.010). The riskiness is higher when the autonomous agents intervened, drivers intervened (rising from 4.47 to 4.65), and the results were severe. The responsibility of drivers is not significantly correlated with any variables. The responsibility of cars is marginally correlated with driver intervention (F=3.829, p=0.063), rising from 4.30 to 5.24. 

\subsubsection{Open-ended Questions on the Emergent Events and Suggestions}
Participants did not identify a large proportion of possible emergent events. 61 out of 189 surveys thought there would be micro accidents, less than one-third. Participants watching original videos proposed improvement suggestions. These suggestions were analyzed using affinity diagram. 139 out of 232 surveys proposed suggestions, among which 43 focused on the driver, 88 focused on cars, and eight proposed that autonomous agents are not trustworthy. 

\section{Discussion}
With the development of automated driving techniques, studies on automated driving has also extended from safety critical and fatal events such as takeover and crash analysis, to broad driving scenarios such as interventions and driving experiences. The posterior scenarios cover a wider range of driving contexts and become a heated application domain for human computer interactions. This article pioneers the study of micro accidents, a state that lie between normal driving and safety critical events. As we illustrated in this article, these micro accidents revealed the weakness of autonomous agents and drivers, acted as the origin of hazard accidents, which could thus be described as the subhealth condition of automated driving. Consequently, micro accidents expand rich research opportunities for driver status computation and adaptive warning systems. 
\subsection{Variables Affecting Micro Accidents Probability}
Multiple variables contributed to micro accidents. Road type is correlated with possibility of micro accidents. Urban roads are prone to ViolationD accidents such as collisions. The irregular construction of rural roads is correlated with misrecognition and wrong warnings, and ChangeB accidents. Road congestion affected the performance of autonomous agents in computer vision and induced ObstacleC accidents. The immaturity of the autonomous agents in sign recognition and route planning invoked a variety of accidents. These variables constituted scenarios that automated driving could not cope well, such as complex intersections, rural areas without signs, highway fork, roads under constructions. These results provided reference for locating the weakness of autonomous agents.

It is worth noting that although the performance of autonomous agents is limited, agents effectively intervened in part of these micro accidents. For example, when vehicles encountered an obstacle or drove to an intersection, the agents initiated an emergent deceleration or braking. In LaneA scenarios (e.g., sway and shake of steering wheel), the agents informed the driver with audio cues while drivers removed the risks. 

\subsection{Subjective Perceptions of Micro Accidents}
The crowdsourcing experiment provided a lens on drivers’ perception of micro accidents. Two-thirds of the participants could not recognize possible micro accidents from videos just before the accidents happened, their risk assessments more influenced by the perceived severity of an event than its probability. This aligns with their ignorance of subtle abnormalities while watching videos. However, drivers’ recognition of risk and intervention are essential to prevent severe accidents. Which risk tends to be ignored warrants further exploration.

Participants’ perceptions are also strongly affected by autonomous agents. The report of computer vision problems increased participants’ ratings of possibilities and the dangerousness of driving scenarios. Participants showed a tendency to rely on autonomous agents to cope with micro accidents. Only a minor emphasized the importance of keeping the driver in the assisted driving state at all times. Therefore, besides an optimization of algorithms, raising people’s awareness during automated driving also have high priority.

\subsection{Insights Derived from the Perspective of Micro Accidents}
This study validated some of the acknowledged results, such as the importance of drivers’ intervention in automated driving, while revealing some new themes. Firstly, drivers’ risk assessments are skewed by the reports from autonomous agents or lack thereof. Therefore, when the agents struggled to handle road planning at forks and did not give reminders, this degraded drivers’ perception of riskiness. Additionally, drivers tend to judge the riskiness by referring to the dangerousness of accidents rather than the possibility of accidents, they thus might ignore and did not intervene when the results seemed less dangerous. This emphasizes the need for new warning modes. Moreover, drivers tended to expect an improvement in autonomous agents rather than reflecting on and improving their own behaviors. This might be the reason for the repetitive occurrence of micro accidents in automated driving, and calls for the training of drivers towards correct perception. 

\subsection{Opportunities for Warning Strategies and Feedback Demos}
The preceding analyses suggests development of adaptive feedbacks. First, we advocate the dynamic warning strategies. As drivers over rely on autonomous agents while perceiving riskiness, current post event reminders are insufficient. Researchers may develop mild coping strategies that lift small problems such as swaying, and build a constantly warning strategy when encountering complex and highly dynamic scenes.

Second, develop specific knowledge pack for automated driving. This would explain the capabilities and limitations of autonomous systems across varied scenarios, thereby empowering drivers with the knowledge of the uncertainty of autonomous agents. Alerts could be activated when vehicles encounter such routes and zones.
\subsection{Limitation of the Study}
It is worth noting that the video we collected is limited. Although this collection involved varied micro-accidents that are hard to find through other methods, it lacks the state of drivers, drivers’ reactions after the accidents, and the responses of surrounding vehicles. There is also potential bias because drivers may tend to upload videos that display the fault of autonomous agents over human error, potentially affecting the SHAP value. Hence, comprehensive automated driving studies that contains whole data of autonomous agents, drivers, and environments still warrants further study.

\section{Conclusion}
This study focused on key variables that affect micro accidents in automated driving. Specifically, we collected a broad range of videos and analyzed using XGBoost and SHAP. A crowdsourcing experiment complemented human perceptions of such micro accidents. Our findings revealed scenarios with which automated driving could not cope well, located key variables that affected the performance of automated driving, and explained drivers’ responses to these micro accidents. This study lay the foundation for refinement of autonomous agents and provide inspiration for interaction design in risky scenarios. Future studies would broaden data sources and incorporate drivers’ mental states for a comprehensive comprehension.




\bibliographystyle{elsarticle-num-names} 
\bibliography{ref-extracts}

\end{document}